\documentstyle[editedvolume,namedreferences]{crckapb} 

\input{epsf}
\newcommand{\EQ}{\begin{equation}}
\newcommand{\EN}{\end{equation}}
\newcommand{\EQA}{\begin{eqnarray}}
\newcommand{\ENA}{\end{eqnarray}}

\newcommand{\Eq}[1]{Eq.~(\ref{#1})}

\newcommand{\Fig}[1]{Figure~\ref{#1}}

\newcommand{\bra}[1]{\langle #1\rangle}

\newcommand{\meanAA}{\overline{\mbox{\boldmath $A$}}}
\newcommand{\meanBB}{\overline{\mbox{\boldmath $B$}}}
\newcommand{\meanJJ}{\overline{\mbox{\boldmath $J$}}}

\newcommand{\yyy}{\hat{\mbox{\boldmath $y$}} {}}

\newcommand{\xx}{\mbox{\boldmath $x$} {}}

\newcommand{\uu}{\mbox{\boldmath $u$} {}}

\newcommand{\bb}{\mbox{\boldmath $b$} {}}

\newcommand{\BB}{\mbox{\boldmath $B$} {}}

\newcommand{\AAA}{\mbox{\boldmath $A$} {}}

\newcommand{\jj}{\mbox{\boldmath $j$} {}}
\newcommand{\JJ}{\mbox{\boldmath $J$} {}}

\newcommand{\eee}{\mbox{\boldmath $e$} {}}

\newcommand{\ff}{\mbox{\boldmath $f$} {}}

\newcommand{\kk}{\mbox{\boldmath $k$} {}}

\newcommand{\nab}{\mbox{\boldmath $\nabla$} {}}

\newcommand{\oo}{\mbox{\boldmath $\omega$} {}}

\newcommand{\DD}{{\rm D} {}}

\newcommand{\dd}{{\rm d} {}}

\newcommand{\yana}[5]{ (#1) #5. {\em Astron. Astrophys. }{\bf #2}, #3--#4}

\newcommand{\ynat}[5]{ (#1) #5. {\em Nature }{\bf #2}, #3--#4}

\newcommand{\yjgr}[5]{ (#1) #5. {\em J. Geophys. Res. }{\bf #2}, #3--#4}
\newcommand{\yapj}[5]{ (#1) #5. {\em Astrophys. J. }{\bf #2}, #3--#4}

\newcommand{\ybook}[3]{ (#1) {\em #2}. #3}

\newcommand{\smn}[2]{  (#1) #2, {\em Mon.\ Not.\ Roy. Astron. Soc.} (submitted)}

\newcommand{\papj}[2]{ (#1)``#2,'' {\em Astrophys. J.} (in press)}

\newcommand{\ea}{{\em et al.\ }}

\def\onethird{{\textstyle{1\over3}}}

\begin{opening}
\title{Sheared helical turbulence and the helicity constraint in large-scale
dynamos}

\author{Alberto Bigazzi}
\institute{Department of Mathematics, Politecnico di Milano,
Piazza Leonardo da Vinci 32, I-20133 Milano, Italy}
\author{Axel Brandenburg}
\institute{Nordita, Blegdamsvej 17, DK-2100 Copenhagen \O, Denmark\\
Mathematics Department, Univ. of Newcastle, NE1 7RU, UK}
\author{Kandaswamy Subramanian}
\institute{National Centre for Radio Astrophysics - TIFR, Poona University Campus,
Ganeshkhind, Pune 411 007, India}

\end{opening}

\runningtitle{Turbulent dynamos with shear}

\begin{document}

\begin{abstract}
The effect of shear on the growth of large scale magnetic fields
in helical turbulence is investigated.
The resulting large-scale magnetic field is also helical and continues
to evolve, after saturation of the small scale field, on a slow resistive 
time scale. This is a consequence of magnetic helicity conservation.
Because of shear, the time scale needed to reach an equipartition-strength 
large scale field is shortened proportionally to the ratio of the
resulting toroidal to poloidal large scale fields.
\end{abstract}

\section{Introduction}

Magnetic helicity is conserved in ideal MHD. 
In non-ideal situations, when magnetic diffusivity is non-vanishing, 
it can only evolve  on a long
time scale governed by microscopic magnetic diffusivity.
This is true 
in periodic or unbounded systems or in systems with perfectly
conducting boundaries, where no flux of magnetic helicity through
the boundaries is allowed. 
In systems with open boundaries, magnetic helicity can leak out 
and the evolution in time can thus be different. 

The importance of magnetic helicity conservation for the evolution of
large-scale magnetic fields in astrophysics has been recently discussed
(Blackman \& Field 2000, Brandenburg 2000, Brandenburg \& Subramanian 2000,
 Kleeorin \ea 2000).  
Large-scale, helical
magnetic fields where the outer scale of
turbulent motions is much smaller than the scale of these fields,
are observed in stars and galaxies. 
For the Sun significant amounts
of magnetic helicity are observed at the solar surface (Berger \&
Ruzmaikin, 2000)

Boundary conditions are important in determining the overall dynamics 
of the large-scale field  (Blackman \& Field, 2000).
In the case of a periodic homogeneous isotropic medium with no externally
imposed magnetic field, recent numerical studies  (Brandenburg 2000, hereafter 
referred to as B2000)  show that, because of magnetic helicity conservation, 
the large scale magnetic
field can only grow to its final (super-) equipartition field strength on
a resistive time scale, which is usually many orders of magnitude longer
than the dynamical time-scale determined by the turbulent eddy turnover time. 

Besides allowing for a flux of magnetic helicity 
through the boundaries by imposing different boundary conditions, 
another way to allow for faster growth of the field 
is  by means of shear which 
can amplify an existing field without changing its  magnetic helicity. 
A regenerative mechanism for the cross-stream (poloidal) component of
the field is also needed, because otherwise 
 the sheared (toroidal) field would eventually decay  (e.g.\ Moffatt 1978,
 Krause \& R\"adler 1980).
 Indeed 
a number of working dynamos which have both open boundaries
and shear have been proposed (e.g., Glatzmaier \& Roberts 1995, Brandenburg \ea 1995),
but those models are rather complex and use sub-grid scale modelling, thus
making it difficult to evaluate the role of magnetic helicity conservation. 

Here we study the effect that shear alone can have on the dynamics of the
large scale field, while keeping the system periodic. 
We find that the evolution of the large scale field is compatible with a mean-field 
model where the geometrical mean of the large-scale poloidal and toroidal
fields evolves on a resistive time-scale. 
It is thus possible to have a larger toroidal field at
the expense of the poloidal one without violating the helicity constraint. 
Equivalently, equipartition strength
large
scale fields can be attained in times shorter by the ratio of the resulting
toroidal to poloidal field strength.

\section{Equations and setup}
\label{Snumer}

The same set of MHD equation for an isothermal compressible
gas as in B2000 is considered.  
The external forcing function $\ff$ incorporates both the helical driving at
intermediate scale $k=5$ and the shear at $k=1$. 

\EQ
{\DD\ln\rho\over\DD t}=-\nab\cdot\uu,
\EN
\EQ
{\DD\uu\over\DD t}=-c_{\rm s}^2\nab\ln\rho+{\JJ\times\BB\over\rho}
+{\mu\over\rho}(\nabla^2\uu+\onethird\nab\nab\cdot\uu)+\ff,
\label{dudt}
\EN
\EQ
{\partial\AAA\over\partial t}=\uu\times\BB
-\eta\mu_0\JJ,
\label{dAdt}
\EN
where ${\rm D}/{\rm D}t=\partial/\partial t+\uu\cdot\nab$ is the
advective derivative, $\uu$ is the velocity, $\rho$ is the density,
$\BB=\nab\times\AAA$ is the magnetic field, $\AAA$ is its vector
potential, and $\JJ=\nab\times\BB/\mu_0$ is the current density. The
forcing function $\ff$ takes the form 
\(
\ff=\ff_{\rm turb}+\ff_{\rm shear},
\)
where
\EQ
\ff_{\rm shear}=C_{\rm shear}{\mu\over\rho}\,\yyy\sin x
\EN
balances the viscous stress once a sinusoidal shear flow has been
established, and
\EQ
\ff_{\rm turb}=N\mbox{Re}\{\ff_{\kk(t)}\exp[i\kk(t)\cdot\xx+i\phi(t)]\},
\EN
is the small scale helical forcing with
\EQ
\ff_{\kk}={\kk\times(\kk\times\eee)-i|\kk|(\kk\times\eee)
\over2\kk^2\sqrt{1-(\kk\cdot\eee)^2/\kk^2}},
\EN
where $\eee$ is an arbitrary unit vector needed in order to generate
a vector $\kk\times\eee$ that is perpendicular to $\kk$, $\phi(t)$
is a random phase, and $N=f_0 c_{\rm s}(kc_{\rm s}/\delta t)^{1/2}$,
where $f_0$ is a non-dimensional factor, $k=|\kk|$, and $\delta t$ is the
length of the time step. 
As in B2000 we choose the forcing wavenumbers such that 
$4.5<|\kk|<5.5$. At each time step one of the 350 possible
vectors is randomly chosen. 

The equations are made non-dimensional with the choice $c_{\rm s}=k_1=\rho_0=\mu_0=1$,
where $c_{\rm s}$ is the sound speed, $k_1$ is the smallest wavenumber in the
box (so its size is $2\pi$), $\rho_0$ is the mean density (which is
conserved), and $\mu_0$ is the vacuum permeability.
The computational mesh is $120^3$ grid-points.
Sixth order finite differences are used for spatial derivatives.

We consider the case when  shear is  strong compared to turbulence, but still subsonic. 
We choose for the shear parameter $C_{\rm shear}=1$ and for the
amplitude of the random forcing $f_0=0.01$. 
The resulting rms velocities in the meridional ($xz$) plane are around 0.015
and the toroidal rms velocities around 0.6.

The magnetic Prandtl number is ten for the simulations considered here, i.e.\
$\mu/(\rho_0\eta)=10$, and $\eta=5\times10^{-4}$. If calculated with respect to
the box size  ($=2\pi$), the Reynolds numbers for poloidal and
toroidal velocities are $R_{\rm m}^{\rm pol}=190$ and $R_{\rm m}^{\rm
tor}=7500$, respectively. By poloidal and toroidal components we mean 
those in the $xz$-plane and the $y$-direction, respectively.
Based,
instead, on the forcing scale, the poloidal magnetic Reynolds number is
only about 40, and the kinetic Reynolds number
is only 4, which is not enough to allow for
a proper inertial range. The turnover time based on the forcing scale
and the poloidal rms velocity is $\tau=70$. 

\section{Time evolution of the field}
\begin{figure}\epsfbox{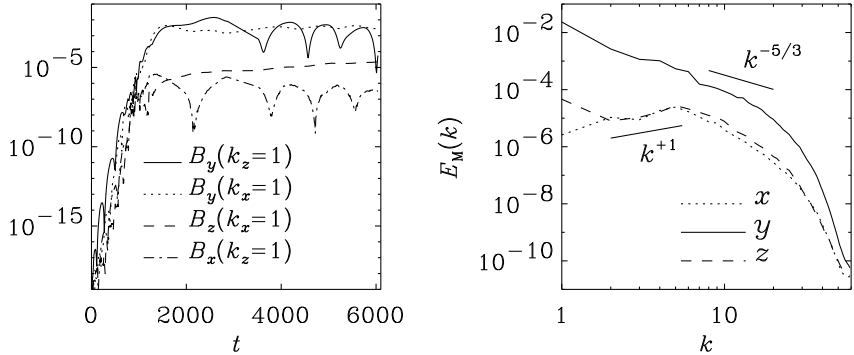}\caption[]{
Evolution of the power, $|\hat{B}_i(k_j)|^2$, of a few selected
Fourier modes (left panel). After $t=1700$, most of the power is in the mode
$|\hat{B}_y(k_z)|^2$, i.e.\ in the toroidal field component with
variation in the $z$-direction. The three-dimensional power spectrum
of the three field components is shown on the right.
$120^3$ mesh-points, $t=5000$.
}\label{Fpbmk1+pspe3d}\end{figure}

A strong dynamo amplifies an initially weak random seed 
magnetic field exponentially  on a dynamical time-scale up to equipartition.
In \Fig{Fpbmk1+pspe3d} we plot the evolution of the
power, $|\hat{B}_i(k_j)|^2$, in a few selected modes.
After
$t=1700$, most of the power is in the mode $|\hat{B}_y(k_z)|^2$, i.e.\
the toroidal field component with variation in the $z$-direction. The
ratio of toroidal to poloidal field energies are around $10^4$, so $B_{\rm
tor}/B_{\rm pol}\approx100$.

A three dimensional power spectrum of the field components in \Fig{Fpbmk1+pspe3d} shows the
different behaviour of the poloidal and toroidal components. 
The dominating toroidal field has a $k^{-5/3}$-like spectrum  
from the largest scale to the dissipative cut-off. 
Poloidal fields, instead, are noisy and possess significant power near $k=5$. 
The poloidal field
saturates earlier than the toroidal, which 
is by then already dominated by large scales.

\epsfxsize=12.5cm

\begin{figure}\epsfbox{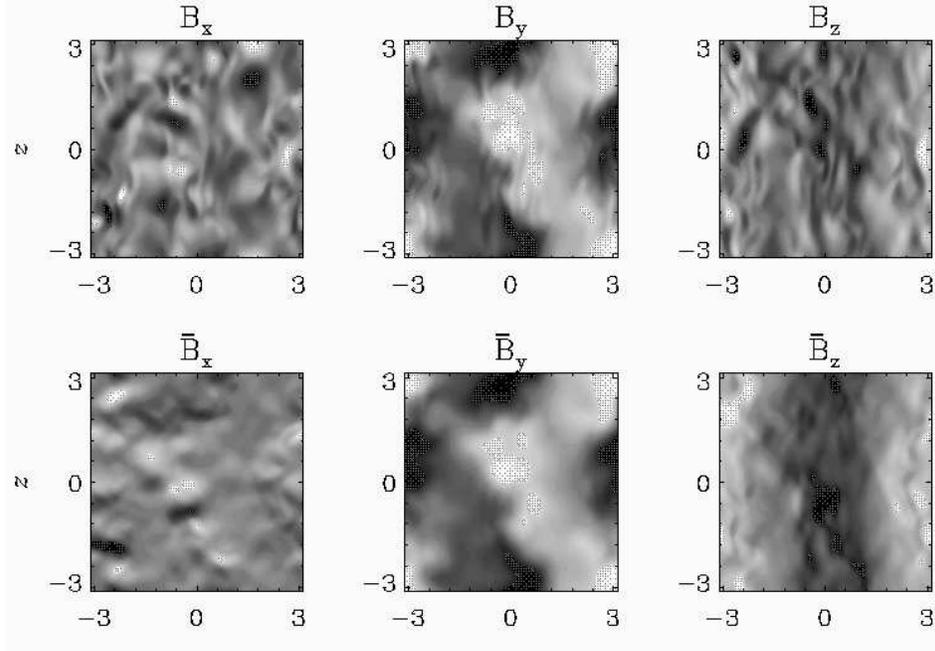}\caption[]{
Images of the three components of $\BB$ in an arbitrarily chosen $xz$ plane
(first row), compared with the $y$-averaged fields (second row).
$120^3$ mesh-points, $t=5000$.
}\label{Fpbmer}\end{figure}

Longitudinal cross-sections  show that the small scale contributions to the poloidal field result from variations
in the toroidal direction. 
Whilst the toroidal field
is relatively coherent in the toroidal direction, the poloidal field
components are much less coherent and show significant fluctuations in
the $y$-direction.
We thus define mean fields $\meanBB$ to be the y-averaged fields. As can be
seen from  \Fig{Fpbmer} this is compatible with the definition 
of the mean field as the large-scale Fourier expansion. 
Two-dimensional power-spectra of this averaged field show that poloidal mean
field components also gain significant
power at the largest scale (i.e.\ at $\kk^2<2$) at later times.

\epsfxsize=12.5cm

\section{Helicity constraint and the mean magnetic field}

In an unbounded or periodic system the magnetic helicity,
$\bra{\AAA\cdot\BB}$, evolves according to 
\EQ
{\dd\over\dd t}\bra{\AAA\cdot\BB}=-2\eta\bra{\JJ\cdot\BB}.
\label{helconstr}
\EN
Taking into account the spectral properties of the above quantities, we may separate large-scale and small-scale contributions and write
\EQ
\bra{\meanJJ\cdot\meanBB}/k_1\approx\mp B_{\rm tor}\,B_{\rm pol}\approx
k_1\bra{\meanAA\cdot\meanBB}.
\label{meanJB}
\EN
The expression above would for instance be true for a field of the form
\EQ
\meanBB=\pmatrix{B_{\rm pol}\,\cos(k_1 z+\varphi_x)\cr B_{\rm tor}\,
\sin(k_1 z+\varphi_y)\cr 0},
\label{BtorBpol}
\EN
where we have allowed for an
additional phase shift between the two components (relative to the
already existing $\pi/2$ phase shift), $\varphi_y-\varphi_x$,
but such a phase shift turned
out to be small in our case. Furthermore, an additional
$x$-dependence of the mean field, which is natural due
to the $x$-dependence of the imposed shear profile could be accounted for. 
However, for the
following argument all we need is relation
(\ref{meanJB}).
The amplitudes  $B_{\rm pol}$ and $B_{\rm tor}$ 
can be calculated as
\EQ
\label{BtorBpolRun}
B_{\rm tor}\equiv\bra{\meanBB_y^2}^{1/2},\quad
B_{\rm pol}\equiv\bra{\meanBB_x^2+\meanBB_z^2}^{1/2}.
\EN
Brackets denote here volume averaging while an overbar indicates an
average over the toroidal direction ($y$).
The upper sign applies to the present case where the kinetic
helicity is positive (representative of the southern hemisphere), and the
approximation becomes exact if \Eq{BtorBpol} is valid.

Following B2000, in the steady case $\bra{\AAA\cdot\BB}=\mbox{const}$,
see \Eq{helconstr}, and so the r.h.s.\ of \Eq{helconstr} must vanish,
i.e.\ $\bra{\JJ\cdot\BB}=0$, which can only be consistent with \Eq{meanJB}
if there is a small scale component, $\bra{\jj\cdot\bb}$, whose sign is
opposite to that of $\bra{\meanJJ\cdot\meanBB}$. Hence we write
\EQ
\bra{\JJ\cdot\BB}=\bra{\meanJJ\cdot\meanBB}+\bra{\jj\cdot\bb}\approx0.
\EN
This yields, analogously to B2000,
\EQ
-{\dd\over\dd t}\left(B_{\rm tor}\,B_{\rm pol}\right)
=+2\eta k_1^2\left(B_{\rm tor}\,B_{\rm pol}\right)
-2\eta k_1|\bra{\jj\cdot\bb}|,
\label{approxB_ode}
\EN
which yields the solution
\EQ
B_{\rm tor}\,B_{\rm pol}
=\epsilon_0 B_{\rm eq}^2\left[1-e^{-2\eta k_1^2(t-t_{\rm s})}\right],
\label{approxB}
\EN
where $\epsilon_0=|\bra{\jj\cdot\bb}|/(k_1 B_{\rm eq}^2)$ is a
prefactor, $B_{\rm eq}$ is the equipartition field strength with
$B_{\rm eq}^2=\mu_0\bra{\rho\uu^2}$, and $t_{\rm s}$ is the time when
the small scale field has saturated which is when \Eq{approxB_ode}
becomes applicable. All this is equivalent to B2000, except that
$\bra{\meanBB^2}$ is now replaced by the product $B_{\rm tor}\,B_{\rm
pol}$. The significance of this expression is that large toroidal fields
are now possible if the poloidal field is weak.

\epsfxsize=12.5cm\begin{figure}\epsfbox{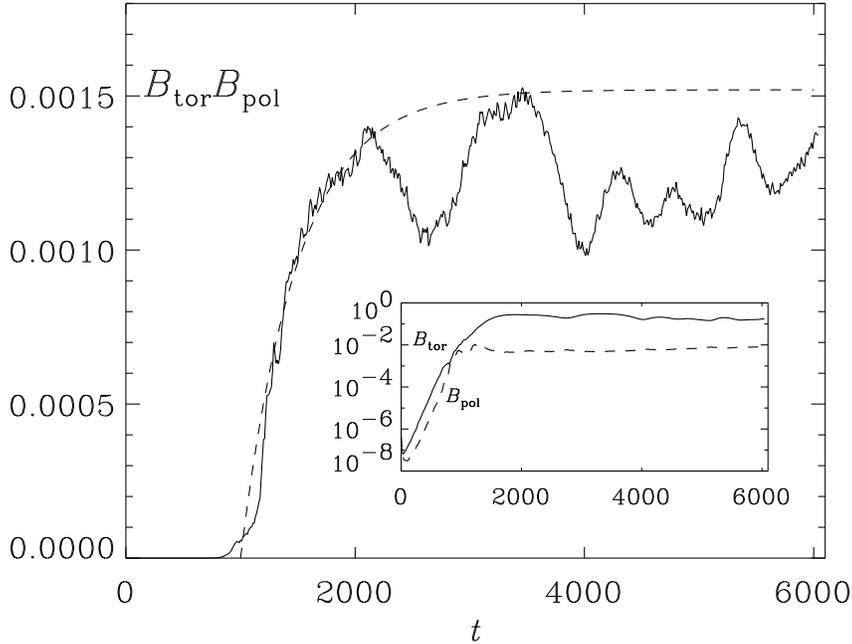}\caption[]{
Growth of the product of poloidal and toroidal magnetic fields on
a linear scale. The inset shows separately the evolution of
poloidal and toroidal fields on a logarithmic scale.
For the fit we have used $k_1^2=2$ and $\epsilon_0=3.8$.
}\label{Fpbm1}\end{figure}

In \Fig{Fpbm1} we show the evolution of the product $B_{\rm tor}\,B_{\rm pol}$
as defined in (\ref{BtorBpolRun}) and compare with \Eq{approxB}.
There are different stages; for $1200<t<2200$ and
$3000<t<3700$ the effective value of $k_1^2$ is $2$ (because there
are contributions from $k_x=1$ and $k_z=1$; see \Fig{Fpbmk1+pspe3d}), whilst at
other times ($2500<t<2800$ and $t>4000$) the contribution from $k_x=1$
(for $2500<t<2800$) or $k_z=1$ (for $t>4000$) has become subdominant
and we have effectively $k_1^2=1$. This is consistent with the change
of field structure discussed in the previous section: for $2000<t<3000$
and around $t=4000$ the $B_y(k_x=1)$ mode is less powerful than the
$B_y(k_z=1)$ mode.

\section{Conclusions}

The effects of the helicity
constraint can clearly be identified in in our system  even though much of the field
amplification results from the shearing of a poloidal field. 
The constraint on
the geometrical mean of the energies in the poloidal and toroidal field
components is evident from \Fig{Fpbm1}. The fit shows that the prefactor
$\epsilon_0$ is about 3.8. Theoretically one may estimate $\epsilon_0$,
which is proportional to $|\bra{\jj\cdot\bb}|$, by estimating
$|\bra{\jj\cdot\bb}|\approx\rho_0|\bra{\oo\cdot\uu}|\approx
k_{\rm f}\rho_0\bra{\uu^2} \approx k_{\rm f}B_{\rm eq}^2$. Since
$\epsilon_0=|\bra{\jj\cdot\bb}|/(k_1 B_{\rm eq}^2)$ this yields
$\epsilon_0\approx k_{\rm f}|/k_1=5$, in good agreement with
the simulation.

Power spectra of the poloidal field show that most of the power
is in the small scales, making the use of averages at first glance questionable. However,
once the field is averaged over the toroidal direction the resulting
poloidal field is governed by large scale patterns (the slope of the
spectrum is steeper than $k^{-1}$, which is the critical slope for
equipartition). The presence even of a weak poloidal field is crucial
for understanding the resulting large scale field generation in the
framework of an $\alpha\Omega$ dynamo.

In another paper (Brandenburg \ea 2000) we have
elaborated further on the similarity between the present simulations and
$\alpha\Omega$ dynamos. In particular, we have discussed anisotropic
turbulent magnetic diffusivities as a possible explanation for the
difference between the resistive growth time of the field on the one hand
and a shorter cycle period seen in the simulation on the other. With
just one simulations so far it is impossible to verify any scaling,
but it is worth mentioning that the present cycle time of around 1000
time units is close to the geometrical mean of resistive and dynamical
timescales. Nevertheless, one must not forget that the real sun does
have open boundaries, and it is now important to understand their role
on the magnetic helicity constraint.

\section*{Acknowledgments}
ABi and KS thank Nordita for hospitality during the course of this
work. Use of the PPARC supported supercomputers in St Andrews and
Leicester is acknowledged.

\end{document}